# Democracy Models and Civic Technologies: Tensions, Trilemmas, and Trade-offs

Marta Poblet[1] and Enric Plaza[2]

[1] RMIT University, Melbourne, VIC 3000, Australia
[2] Artificial Intelligence Institute, Spanish Research Council, Bellaterra 08193 Spain
marta.pobletbalcell@rmit.edu.au
enric@iiia.csic.es

**Abstract.** This paper aims at connecting democratic theory with civic technologies in order to highlight the links between some theoretical tensions and trilemmas and design trade-offs. First, it reviews some tensions and 'trilemmas' raised by political philosophers and democratic theorists. Second, it considers both the role and the limitations of civic technologies in mitigating these tensions and trilemmas. Third, it proposes to adopt a meso-level approach, in between the macro-level of democratic theories and the micro-level of tools, to situate the interplay between people, digital technologies, and data.

## 1 Introduction

Over the last two decades, digital technologies have opened up new paths for civic engagement and political participation. Hundreds of websites, portals, platforms and mobile apps enable citizens across the globe to organise campaigns, vote initiatives and sign petitions, monitor their representatives and track parliamentary activity, propose ideas and draft legislation or constitutions. Governments at different levels adopt digital technologies to develop 'open government' and 'open data' strategies to promote citizens' participation and increase transparency. Crowdsourcing is now a pervasive method to collect data, information, ideas, and legislative proposals. A growing literature based on case studies and empirical testing provides the basis for further refinement of methods: e.g. smart crowdsourcing (Noveck 2015), expert crowdsourcing (Kim et al. 2014, Griffith et al. 2017), microtasking (Luz et al. 2015).

The exploration of new technologies and methods to harness the potential of crowdsourcing for civic action and politics, nevertheless, contrasts with the scarce attention given to the underlying assumptions about democracy, participation, equality, representation, and citizenship. Surprisingly enough, there has been little dialogue between theorists of democracy and citizenship, on the one hand, and digital technologists, information systems and AI experts, on the other, on how civic technologies may redefine our current notions of democracy, participation, equality, representation, and citizenship.

Our goal in this paper is threefold. First, we aim to induce a discussion on how to reinterpret some of these notions by reviewing some tensions and 'trilemmas' raised by political philosophers and democratic theorists. Second, we consider both the role and the limitations of civic technologies in mitigating these tensions and trilemmas. Third, we propose to adopt a meso-level approach, in between the macro-level of democratic theories and the micro-level of tools, to situate the interplay between people, digital technologies, and data. As different groups in different social contexts use digital tools and data differently, it is at this meso level that we can elucidate the trade-offs with the notions of the trilemmas. We conceptualise the meso-level as the institutional level, for the notion of institution will give us a framework to analyse the use of technology in a given social context.

## 2 Some Tensions and 'Trilemmas' in Democratic Theory

### 2.1 A Condorcetian Reading of Representation

The tensions between key concepts in democratic theory, notably sovereignty, representation, participation, equality, and citizenship have long been debated. In her work on representative democracy, Nadia Urbinati has noted that both Montesquieu and Rousseau were 'the first theorists to argue (for divergent reasons) that an unsolvable tension exists between democracy, sovereignty, and representation' (2004: 54). More specifically:

> Montesquieu separated representation from democracy, and Rousseau representation from sovereignty. Montesquieu argued that a state where the people delegated their 'right of sovereignty' could not be democratic and must be classified as a species of mixed government and in fact an aristocracy. Rousseau saw such a state as non-political from the start and illegitimate because the people lost their political liberty along with the power to vote on legislation directly: unless all citizens were lawmakers, there were no citizens at all. In both cases, democracy and sovereignty excluded representation (idem).

Urbinati argues that this exclusion remains implicit within contemporary theories of representative government for which "from a theoretical point of view, a 'represented democracy', although technically feasible, is an oxymoron, while direct democracy, although the norm, is impractical" (2004: 55). Yet, Urbinati denies this incompatibility to be the only legacy of 18th century's political philosophy when it comes to the idea of representation.[1] In supporting her claim for a 'democratic under-standing of representation' she draws on Condorcet's *Plan de Constitution* submitted to the French National Assembly in 1793. Condorcet's proposal, eventually rejected by both his fellow Girondins and the Jacobins, contains what Urbinati describes as 'an institutional order that is one of the most democratically advanced and imaginative Europe has produced in the last two centuries' (2004: 56):

> Condorcet's constitution designed a political order that was horizontal and acephalous (parliamentary, not presidential) and rigorously based on the centrality of the legislative power, a power held by a multiplicity of actors and performed in multiple times and within a plurality of spaces. The function of legislation was performed within assemblies – elected assembly and assemblies of the citizens (*assemblées primaires*) – and was held by the representatives along with (not instead of) the citizens who 'enjoyed' both the electoral right and the right to revoke or censure the laws (constitutional and ordinary). (2004: 59-60).

[1] 'Rather than a monolithic entity, the theory of representative government formed, since its birth, a complex and pluralistic family whose democratic wing was not the exclusive property of those who advocated for participation against representation.' (Urbinati 2004: 55).

Condorcet, Urbinati notes, reconciles sovereignty, representation, and participation by making 'citizens' participation essential to both the functioning of representative government and the preservation of political liberty' (2004: 60). With a comment that echoes Josiah Ober's vision of the role of citizens in ancient Athens (2008, 2015), Urbinati sees citizen participation in Condorcet's institutional order as a 'source of stability and of innovation', while representation becomes the political device collecting and filtering knowledge for the public interest (2004: 60).

In our contemporary democracies, representation has become an even more intricate subject, even at the local level (Ng *et al.* 2016). Urbinati and Warren argue that the complexity of issues and the multiple, overlapping constituencies involved call for the extension of the meaning of representation to include non-electoral forms 'that are capable of representing latent interests, transnational issues, broad values, and discursive positions' (Urbinati and Warren 2009: 407). Moreover, the Internet has also enabled the emergence of online communities of interest beyond geographical boundaries that have no mechanisms of representation in our political systems (Lloyd 2017).

It is our contention that digital technologies and AI can facilitate the channelling of these multifaceted forms of representation in unique ways. But a second 'trilemma' needs to be addressed before considering these options.

### 2.2 The 'Trilemma' of Democratic Reform

James Fishkin, a leading theorist of deliberative democracy, addresses in one of his papers the key question of how to incorporate public deliberation into constitutional processes (Fishkin 2011). In raising this question he introduces what he refers to as the 'trilemma of democratic reform'. To Fishkin, there are three basic principles internal to the design of democratic institutions: political equality (people's views are counted equally), mass participation (we are all given the opportunity to provide informed consent), and deliberation (we are all given the opportunity to provide opinions and weigh competing arguments).

Fishkin suggests that, under normal conditions, any serious effort to attain any of the two principles inevitably hinders the third, so that we cannot satisfy the three prin-ciples simultaneously. For example, if we pursue a process driven by political equality and mass participation we are unlikely to get deliberation into the picture because the incentives for people to become seriously informed and engaged are very low ('audience democracy'). Likewise, we can satisfy the principles of political equality and deliberation if we choose (by lot or by random sampling) a microcosm of deliberators (e.g. Fishkin's Deliberative Polls). This microcosm may be representative of the broader population from which it has been extracted, but then this population will have no voice in the process and therefore the principle of mass participation will not be fulfilled. Finally, we can have a process with mass participation (to some extent) and deliberation. This is what most of the current online crowd-civic platforms provide, but what we gather in this case is a 'self-selected microcosm of deliberators', highly engaged and yet, far from being representative of the broader population (so we would be violating the principle of political equality). Tanja Aitamurto *et al.* (2014) have also highlighted the tension between the norm of equal representation in democracy and the self-selection bias of crowdsourcing, suggesting that 'crowdsourcing shouldn't strive for statistical representativeness of the public, otherwise the virtues of crowdsourcing would be compromised and its benefits in crowd work would not be achieved.' (2014: 1). Statistical representativeness as a requirement may be a debatable issue, but what is at stake here is the *legitimacy* of crowdsourcing in political practice. We also find a self-selection bias in offline political activity, e.g. in parliamentary elections, where the turnout is usually significantly below 100 per cent of the demos. How self-selection affects legitimacy in a political process is a general issue that political theory needs to address in broader terms. Specifically, if we conceptualize political equality in the classical sense [*isegoria* (equal voice) + *isonomia* (equality of political rights)] self-selection does not necessarily diminish the principle of equality (non-participation is an individual decision).

What should we do if the simultaneous achievement of the three principles is not attainable? Fishkin suggests adopting a pragmatic approach to solve his trilemma. Rather than trying to approximate the ideal, he proposes the design of a second best approach or a proxy (and hence his research program on Deliberative Polling, aiming at both the internal and external validity of the process). Nevertheless, Fishkin acknowledges that this solution may incur a democratic deficit, since the resulting views may not be the actual views of the public (2011: 253). To tackle this issue, he proposes a process with sequential strategies (for example, a convention followed by a deliberative microcosm and, finally, a referendum) that, combined, cover the three principles at different stages.

The remaining issue, nevertheless, is that deliberation does not travel well across those stages. Fishkin illustrates what he terms 'the weak link of deliberation' with the example of the Australian 1999 referendum, where two different deliberative bodies (a convention and a deliberative poll) had previously reached the opposite conclusion (pro-republic) with regard to the proposal of an Australian republic (2011: 253-254). The elaboration of Iceland's Constitution is another recent example of the weak connection between deliberative bodies (in this case, the Constitutional Convention and the Parliament). Fishkin proposes to strengthen this link by organizing a Deliberation Day, where the entire population is convened for one day to engage in deliberation followed by a referendum. To motivate participants, Fishkin estimated that an incentive of $300 per participant would act as an adequate incentive (2011: 258). No matter how well designed, though, the costs of such events could be extremely prohibitive for many countries, especially considering how short-lived they would be. The question that remains open is whether there is a role for technology in mitigating the trilemma.

## 3 Mitigating Democratic Trilemmas

Political philosophy addresses both the tensions and trilemmas in democratic theory and practice with a sophisticated conceptual apparatus. Yet, research on the implications of civic technologies for democracy and democratisation processes is still largely overlooked in both deliberative and epistemic accounts of democracy. This compartmentalisation of knowledge is disadvantageous from both a theoretical and empirical perspective. For example, enabling effective non-electoral forms of representation would require a survey of technology options and 'knowledge of what works and when' (Noveck 2017). Likewise, a better understanding of the underlying principles, models, and concepts of democratic theory would help to inform the design of civic tools and modulate the frequently inflated expectations placed on them.

Digital platforms facilitate the depth and breadth of participation, lowering the barriers to different forms of participation (without precluding offline participation) and improving the 'open access pattern' of a given social order (North *et al*. 2009). They also open up the door to new, meaningful forms of mass deliberation and epistemic outcomes (e.g. Klein 2015, Luz *et al*. 2015, Theocharis and van Deth 2016). To illustrate this point, in Figures 1 and 2 below we compare two models of democracy:

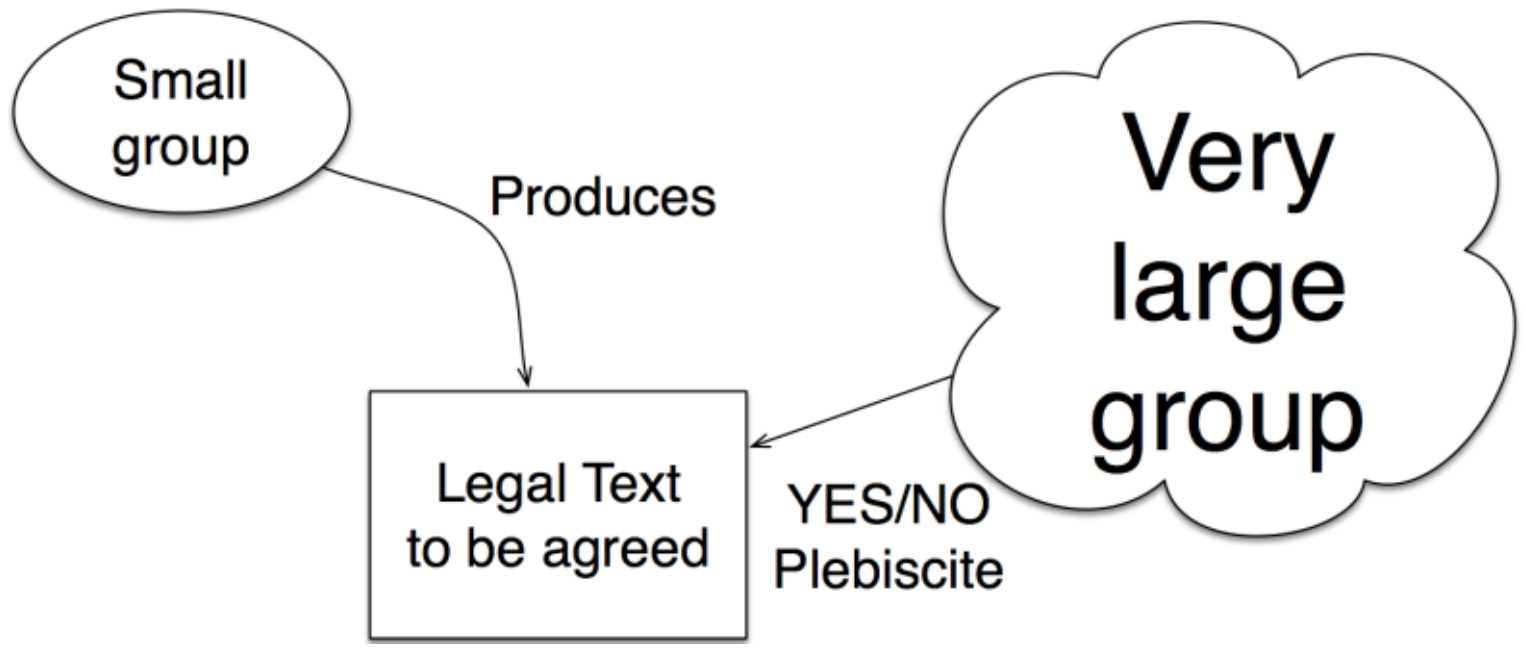

**Fig. 1.** Plebiscitarian model with deliberative

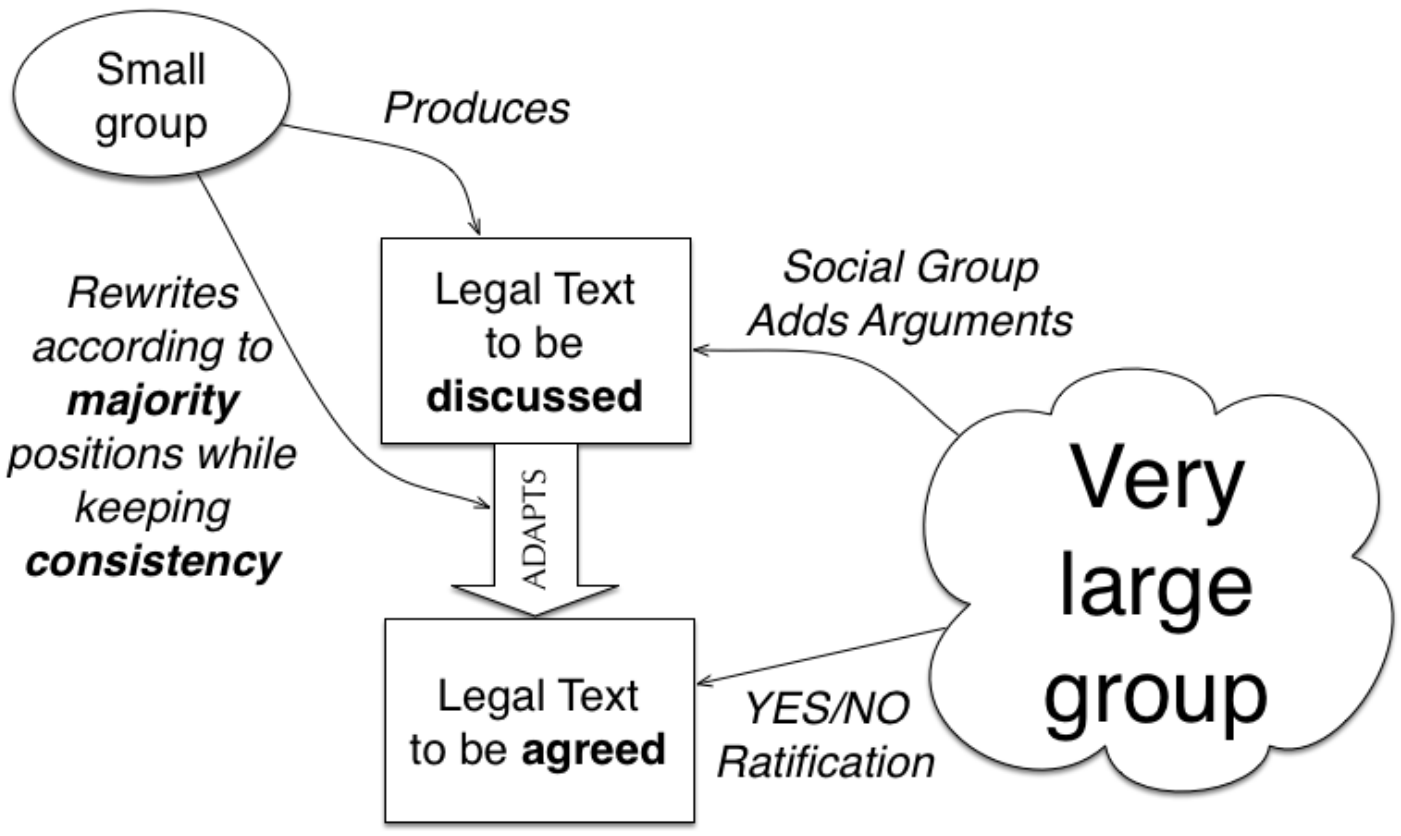

**Fig. 2.** Participatory model with deliberative body

Fig. 1 represents a well-known plebiscitarian model of democracy: a small group (for example, a constitutional convention, a parliamentary commission, etc.) produces a legal text. When the text is ready, a referendum is called and citizens can cast a yes/no vote. This model accounts for the principles of political equality, mass participation, and representation. Yet, deliberation is restricted to the small group, as citizens are left with just an *ex post*, binary option (yes/no). Many constitution making processes in Western democracies have followed this path to date.

Fig. 2 visualises a more complex participatory model to mitigate the trilemma. As in the previous case, a small group of people (either drafted by sortition or appointed by some other entity) is given the task of producing a legal text, but in sequential steps. The group deliberates on a first draft, which is open to the general public for comments and suggestions (typically from a self-selected subset of the electorate). The feedback from this very large group is incorporated in the draft and subsequently adapted to produce, after a number of iterations, the text to be agreed and ratified by the electorate. This participatory model was famously deployed in Iceland in 2011, when the meetings and debates of a Constitutional Council of 25 individuals (drafted from a larger pool of citizens) were made publicly available in the Council website for comment via social media and e-mail. The proposal was approved by a two-thirds majority of the voting population in a referendum in late 2012 but it eventually stalled in parliament (Landemore 2017).

Similarly, this model was adopted in Mexico City. On January 2016, the Mayor of the city obtained approval from the federal parliament to initiate a constitution-making process by appointing a group of 30 experts to discuss and draft a proposal.[2] In order to open up the drafting process to the citizenry, the City Council made publicly available a collaborative editing tool for citizens to provide feedback on the specific topics posted by the drafting group. Moreover, as crowdsourced legal drafting does not typically attract a large number of citizens, this approach was complemented with other participatory strategies, namely a survey and a Change.org campaign to collect petitions relevant to the constitutional text (at the closing date of the process, 280,678 people had supported 129 petitions). The Constitution of Mexico City was finally published on 5 February 2017, although at the time of writing the Supreme Court of Mexico is hearing a number of appeals to the constitutional text from the federal government, two political parties, and other organisations.[3] Strikingly, both the Icelandic and Mexican constitutional drafts came to a standstill as other institutional bodies were involved. We will review this in Section 3.2 below.

[2] https://www.constitucion.cdmx.gob.mx/constitucion-cdmx/#grupo-trabajo

### 3.1 The Technology Caveat

Digital platforms have come a long way when it comes to facilitating legal drafting, crowdsourcing of ideas, or structuring large-scale deliberation, but the tasks of aggregating legal and political knowledge for deliberation and decision making remain onerous. In recent years, a number of advances in AI areas such as text mining, argument detection, extraction, and mapping can be applied to support the activity of very large groups, both to improve self awareness (of what they are co-producing) and facilitate knowledge aggregation. Likewise, both small and large groups can benefit from text mining, semantic languages (e.g. RDF, XML), ontologies, linked data, and machine learning when searching, analysing and reusing legal texts to elaborate new ones. For example, using ConstituteProject,[4] constitution makers can now browse nearly 200 constitutions across the world (tagged with more than 300 topical labels) when drafting their own. Global laws are also accessible to law proponents or drafters with services offered, among others, by the World Legal Information Institute[5] or Global Regulation.[6]

To date, online platforms have focused on improving and facilitating mass participation (or at least to include larger numbers of citizens in a political process). Those efforts have proved useful when supporting the participation of dozens, hundreds or, in some cases, thousands of people contributing to an initiative with arguments or comments. Yet, the issue of effectively enabling large-scale, massive participation (that is, hundreds of thousands or even millions of people) is still unresolved.

It is also important to note here the implicit assumption that correlates higher participation with higher legitimacy. Mexico City, to use our previous example, has almost 9 million inhabitants, but what is the threshold for establishing that a constitution crowdsourced from a negligible percentage of its inhabitants is more legitimate than appointing a group of 30 experts? Can future civic technologies really scale up to *mass* participation in elaborating policies and laws, or can legitimacy only be claimed when the crowds are requested to ratify them? Would it be better to design systems that cater for smaller, decentralised, and distributed (offline and online) citizen as-semblies (thus supporting a renewed version of democratic representation)?[7] While these questions remain open, the answers also depend on political and institutional choices.

[3]http://www.eluniversal.com.mx/articulo/nacion/politica/2017/03/10/corte-admite-impugnaciones-contra-constitucion-cdmx
[4] http://constituteproject.org
[5] http://wordlii.org
[6] http://www.global-regulation.com

## 3.2 The Institutional Caveat

A second caveat when trying to mitigate democratic trilemmas is that deploying civic tools for large-scale participation will not guarantee any real influence on either rule making or policy making. As the examples in Iceland and Mexico show, there is no way to ensure that embedding participatory components into the process—regardless of whether this participation is deliberative or not—will eventually have an impact on decision making and, ultimately, will lead to more bottom-up, inclusive decisions.

Over the last two decades, deliberative democrats have set the conditions, procedures, and standards of deliberative processes. More recently, some of them have adopted a 'systemic' approach where some institutions will achieve some principles while others will achieve others, making the institutional system 'deliberative' as a whole (Mansbridge *et al*. 2012). The focus on procedures and standards has also expanded to include the discussion on whether mini-publics (citizen juries, citizen assemblies, deliberative polls, etc.) and other institutional innovations should have a binding force—aligning the outcomes of deliberation with rule or policy making—or have a mere advisory role (e.g. Lafont 2015). The debate highlights the underlying tensions between participation and deliberation, but it does so from an abstract perspective. Ironically enough, the discussion on the optimal institutional design to coordinate and translate deliberative outputs at the micro level into aligned policy making is not institutionally anchored. Yet, without such anchoring, it is hard to predict in which particular institutional contexts the new designs will either thrive or languish, and which trade-offs will be required. Empirical studies focusing on the institutional level, such as the Utrecht experiment below, may help to shed some light:

> The key feature of this process of political innovation is that citizens were randomly selected to participate, they received remuneration for their participation and they could be regarded as an alternative form of citizen representation. In contrast with many other forms of participation such as citizen panels, the advice was not 'free': local government had committed beforehand to follow this advice and to translate it to an energy policy plan. Our empirical analysis of this case shows that an interplay between idealist and realist logics explains why they are 'accepted' by the institutionalized democratic system." (Meijer *et al*. 2016: 21).

---

[7] We also find examples of this option in Buenos Aires, British Columbia, or Ireland. The Swiss 'semi-direct democracy' model (Cormon 2015) is paradigmatic when combining representation and popular sovereignty at the three levels of governance (federal, cantonal and municipal). Approximately four times a year, voting occurs over various issues: federal popular initiatives (constitutional reforms), policies, and election of representatives. Federal, cantonal and municipal issues are polled simultaneously, and the majority of votes are cast by mail.

An intermediate, meso-level approach to both online and offline innovations would help to elucidate the interactions between people, technology, and data in particular settings. It would also provide a framework of analysis to better understand both the emerging properties (and tensions) of these interactions. We have denominated this approach 'linked democracy' (Poblet *et al*. forthcoming).

## 4 A Proposal for a Meso-level Approach: Some Features

Our proposal consists of analysing political ecosystems where clusters of institutions are distributed throughout with different roles and specialisations, but all connected together. Both the Mexican and Icelandic cases can be analysed through these lenses, as well as, for example, the connected interactions between people, technology and data in a public health ecosystem (e.g. Casanovas *et al*. 2017).

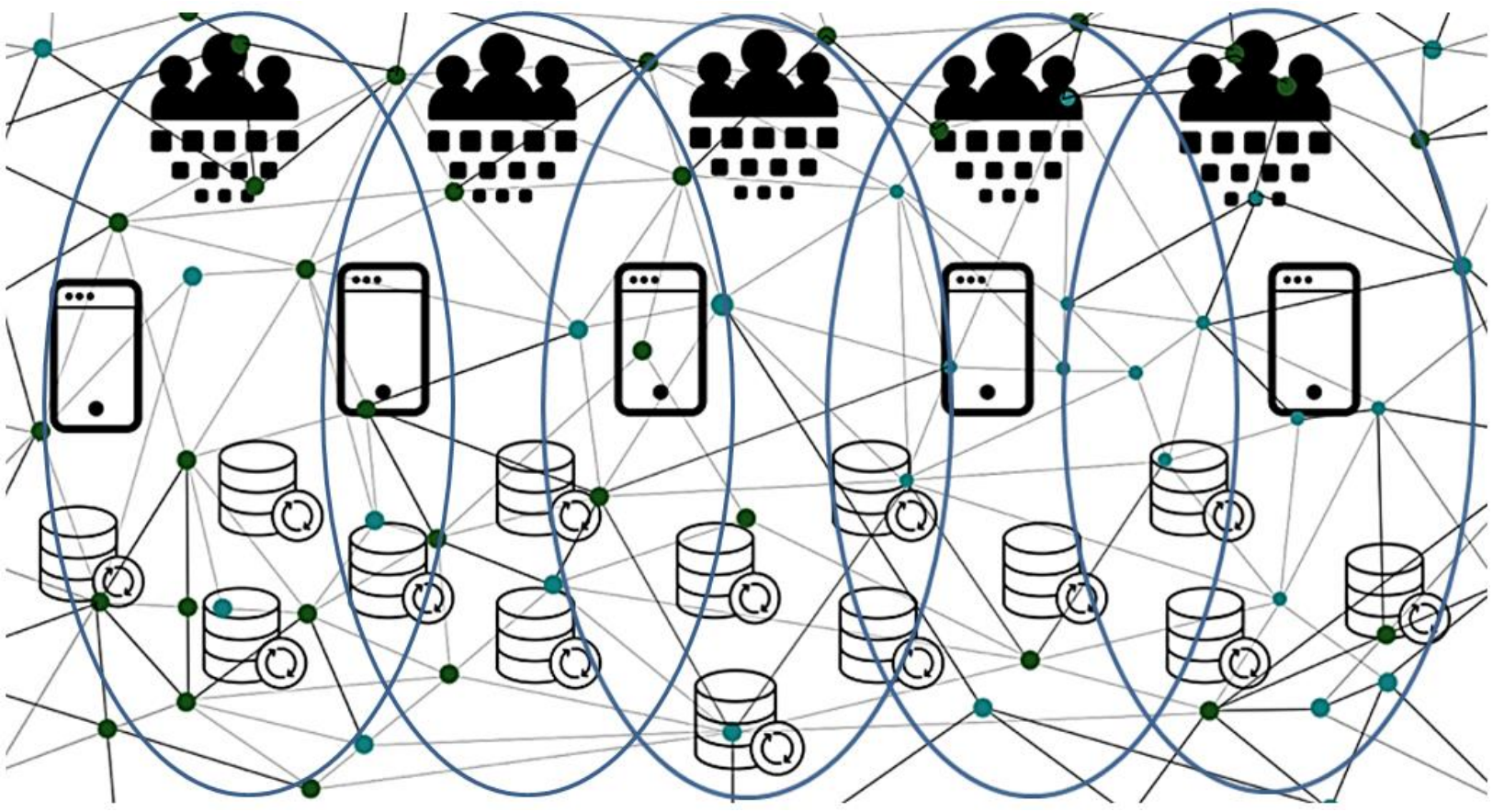

**Fig. 3.** An ecosystem of linked institutions[8]

It is out of the scope of this paper to present a case study embedded in this meso-level approach. Nevertheless, our proposal here includes considering the features that may guide such an analysis from the perspective of a linked democracy theory. Thus, our analysis of political ecosystems will consider them as:

- **Contextual**. Interactions between people, technologies, and data occur at specific settings. People are identifiable individuals or groups, geographically bounded or connected online (or both); technologies include specific devices and tools (platforms, apps, sensors, etc.); data comprises particular datasets with different formats (open data, linked open data, etc.) and licenses of use.

[8] In Figure 3 we use icons from the Noun Project (https://thenounproject.com/): group icon by Gregor Cresnar; data icon by IcoDots; mobile device icon by Vildana.

- **Distributed**. Political ecosystems are distributed networks with multiple nodes (as opposed to centralised or decentralised systems). The level of connectivity is not a given and requires analysis and measurement. Most likely, different political ecosystems will exhibit different connectivity maps—or political 'connectomes', to borrow an emerging concept from the neurosciences (Seung 2013).
- **Open ended**. Political ecosystems will evolve and adapt as the context changes. Stakeholders and their interests are not stable, technologies change rapidly and data has been characterised with the 4 Vs (volume, velocity, variety, and veracity). If any, a theory of linked democracy is a theory of adaptive complex systems.
- **Reusable**. Political ecosystems produce collective knowledge. Both deliberation and epistemic democracy approaches assume the need to find and reuse this knowledge. Ober (2008, 2015) adds to this the dimension of problem solving, in the sense that untapped knowledge can only be 'discovered' in relation to a particular political issue, by making a connection of relevance between that knowledge and the issue at hand. We are interested in discovering how those connections are made and how they can be reused.

## 5 Concluding Remarks

In this paper we have briefly sketched some tensions and trilemmas in democratic theory that are relevant to the topic of designing civic technologies for democracy. Our contention is that technology can provide solutions to these tensions and trilemmas if we embed the issues at stake in a particular institutional meso-level.

Most online platforms focus on facilitating engagement and specially participation. As we have seen, it is not possible to scale up participation by mere technological prowess. Developing technological platforms in the near future will require an integrated approach where trade-offs between political values are explicitly acknowledged and the institutional design of the different components and processes is coherent with contextual constraints and changing environments. Civic values are also critical, and we agree with the perspective of Shannon Vallor (2017) when she states that 'the designs of such platforms have assumed civic virtues as inputs, rather than helping to cultivate them—virtues like integrity, courage, empathy, perspective, benevolence, and respect for truth necessary to fuel any democratic technology, analog or digital'. A theory of linked democracy is proposed to pay attention to these different dimensions.

### References

1. Aitamurto, T., Galli, J.S. and Salminen, J. (2014). Self-selection in crowdsourced democracy: A bug or a feature?http://www.jorgesaldivargalli.com/w3/papers/AitamurtoSG14.pdf
2. Casanovas, P., Mendelson, D. & Poblet, M. (2017). A Linked Democracy approach to regulate health data. *Health and Technology*, doi: 10.1007/s12553-017-0191-5
3. Cormon, P. (2015). Swiss Politics for Complete Beginners (2 ed.), Geneva: Slatkine.
4. Fishkin, J. S. (2011). Deliberative democracy and constitutions. *Social Philosophy and Policy*, 28(01): 242-260.

5. Griffith, M., Spies, N.C., Krysiak, K., McMichael, J.F., Coffman, A.C., Danos, A.M., Ainscough, B.J., Ramirez, C.A., Rieke, D.T., Kujan. L., & Barnell, E.K. (2017) CIViC is a community knowledgebase for expert crowdsourcing the clinical interpretation of variants in cancer. *Nature genetics* 1, 49(2): 170-4.
6. Kim, J., Cheng, J., & Bernstein, M.S. (2014). Ensemble: exploring complementary strengths of leaders and crowds in creative collaboration. In *Proceedings of the 17th ACM conference on Computer Supported Cooperative Work & Social Computing*; 745-755.
7. Lafont, C. (2015). Deliberation, participation, and democratic legitimacy: Should deliberative mini-publics shape public policy? *Journal of Political Philosophy*, *23*(1): 40-63.
8. Landemore, H. (2017). Inclusive constitution making and religious rights: Lessons from the Icelandic experiment. *The Journal of Politics*, *79*(3): 000-000.
9. Lloyd, A. (2017). Disentangling Democracy From Geography. *The Atlantic*. May 2, 2017, https://www.theatlantic.com/technology/archive/2017/05/disentangling-democracy-from-geography/524124/
10. Luz N., Poblet M., Silva N., & Novais P. (2015) Defining human-machine micro-task workflows for constitution making. In: Kamiński B., Kersten G., Szapiro T. (eds) *Outlooks and Insights on Group Decision and Negotiation. GDN 2015*. Lecture Notes in Business Information Processing 218: 333-344.
11. Mansbridge, J., Bohman, J., Chambers, S., Christiano, T., Fung, A., Parkinson, J., & Warren, M. E. (2012). A systemic approach to deliberative democracy. In J. Parkinson and J. Mansbridge (eds.) *Deliberative systems: deliberative democracy at the large scale*. Cambridge University Press: 1-26.
12. Meijer, A., Van der Veer, R., Faber, A, & Penning de Vries, J. (2017). Political innovation as ideal and strategy: the case of aleatoric democracy in the City of Utrecht. *Public Management Review*, 19(1): 20-36.
13. North, D. C., Wallis, J. J., & Weingast, B. R. (2009). *Violence and social orders: a conceptual framework for interpreting recorded human history*. Cambridge University Press.
14. Ober, J. (2008a). *Democracy and knowledge: Innovation and learning in classical Athens*. Princeton University Press.
15. Ober, J. (2015). *The rise and fall of classical Greece*. Princeton University Press.
16. Ng, Y. F., Coghill, K., Thornton-Smith, P., & Poblet, M. (2016). Democratic representation and the property franchise in Australian local government. *Australian Journal of Public Administration*. doi: 10.1111/1467-8500.12217.
17. Noveck, B.S. (2017). Five hacks for digital democracy. *Nature* 544**,** 287–289 (20 April 2017). doi:10.1038/544287a
18. Poblet, M., Casanovas, P., Rodriguez-Doncel, V. (forthcoming, 2017). *Linked Democracy: Foundations, tools, and applications*. Springer Open.
19. Theocharis, Y., & van Deth, J. W. (2016). The continuous expansion of citizen participation: a new taxonomy. *European Political Science Review*: 1-24.
20. Seung, S. (2013). *Connectome: How the brain's wiring makes us who we are*. Boston: Mariner Books.
21. Urbinati, N. (2004). Condorcet's democratic theory of representative government. *European Journal of Political Theory*, 3(1), 53-75.
22. Urbinati, N., & Warren, M. E. (2008). The concept of representation in contemporary democratic theory. *Annual Review of Political Science*, 11: 387-412.
23. Vallor, S. (2017). Lessons From Isaac Asimov's Multivac, *The Atlantic* (May 2, 2017), https://www.theatlantic.com/technology/archive/2017/05/lessons-from-the-multivac/523773/